\documentclass[sigconf]{acmart}
\usepackage{graphicx} 
\renewcommand\footnotetextcopyrightpermission[1]{} 

\PassOptionsToPackage{hyphens}{url}\usepackage{hyperref}
\usepackage[nameinlink, noabbrev, capitalize]{cleveref}
\usepackage[table, svgnames, dvipsnames]{xcolor}

\usepackage{color}

\usepackage{changepage}
\newenvironment{assignment}
  {%
   \par\medskip
   \begin{adjustwidth}{0.075\linewidth}{0.075\linewidth}
   \begingroup
   \itshape
   \setlength{\parindent}{1.5em}
   \setlength{\parskip}{0.5em}
  }
  {%
   \par\endgroup
   \end{adjustwidth}
   \medskip\par
  }

\newtheorem{question}{Research Question}

\author{Ethan Dickey}
\affiliation{
  \institution{Purdue University}
  \streetaddress{Department of Computer Science}
  \city{West Lafayette}
  \state{Indiana}
  \country{USA}
  \postcode{47907}
  }
\email{dickeye@purdue.edu}
\orcid{0009-0007-3706-5253}
\author{Marios Mertzanidis}
\affiliation{
  \institution{Purdue University}
  \streetaddress{Department of Computer Science}
  \city{West Lafayette}
  \state{Indiana}
  \country{USA}
  \postcode{47907}
  }
\email{mmertzan@purdue.edu}
\author{Alexandros Psomas}
\affiliation{
  \institution{Purdue University}
  \streetaddress{Department of Computer Science}
  \city{West Lafayette}
  \state{Indiana}
  \country{USA}
  \postcode{47907}
  }
\email{apsomas@purdue.edu}

\ccsdesc[500]{Computing methodologies~Artificial intelligence}
\ccsdesc[500]{Social and professional topics~Computing education}

\keywords{Generative AI, large language models, computing education, algorithms, experimental design}

\title{Is Solving Better Than Evaluating GenAI Solutions?}

\begin{abstract}
As Generative AI (GenAI) tools become increasingly capable of generating solutions to computing assignments, the computing education community is exploring pedagogical approaches that emphasize solution evaluation, verification, and critique alongside traditional solution generation. However, empirical evidence regarding the impact of such evaluation-centered tasks on student learning remains limited, particularly in upper-division, theory-heavy courses. We conducted a randomized A/B crossover study ($N=220$) in a junior-level algorithms course to investigate how evaluating GenAI-generated solutions compares with traditional problem solving. Across six assignments, student working groups either solved challenging algorithmic problems directly or evaluated often-flawed GenAI-generated solutions to corresponding problems, with group roles reversed midway through the semester.

We found no statistically significant differences between groups on midterm scores, final exam scores, overall course grades, or exam problems structurally aligned with the homework interventions. Students received significantly higher homework scores during periods in which they evaluated GenAI-generated solutions, but this localized advantage on the modified homework items did not translate into downstream summative gains. Survey data further indicated that most students did not report changing their study habits in response to the intervention; however, students who did report adapting their study strategies rated the GenAI-evaluation assignments as significantly more helpful. These findings suggest that GenAI evaluation redistributes student effort from open-ended solution construction toward verification, diagnosis, and judgment, but does not automatically produce stronger conceptual transfer. We conclude that GenAI-evaluation activities can be incorporated into algorithms coursework without broad performance losses, but that meaningful learning gains may require deliberate scaffolding that pushes students beyond simple error diagnosis.
\end{abstract}

\begin{document}

\pagestyle{plain}

 \maketitle

\section{Introduction}

In recent years, advances in Generative Artificial Intelligence (GenAI) and Large Language Models (LLMs) have fundamentally reshaped computing education.  Modern LLMs are capable of generating syntactically correct code and solving many introductory programming and algorithms problems \cite{denny2024computing, finnieansley2022robotscoming,finnie2023my, reeves2023evaluating,savelka2023thrilled}. As these tools become increasingly accessible, students are no longer merely learning about artificial intelligence; they are actively incorporating AI systems such as ChatGPT and GitHub Copilot into their day-to-day programming workflows \cite{bird2022taking}. This rapid adoption has sparked widespread discussion regarding how computing curricula, pedagogical practices, and assessment methods should adapt in response to these technologies \cite{prather2023robotshere, denny2023Chat, qiao2026systematic}. 

The emergence of GenAI presents both substantial opportunities and significant risks for computing education. Prior work has shown that LLMs can provide detailed code explanations, help debug errors, deliver personalized assistance, and offer interactive support that may improve student engagement and productivity \cite{leinonen2023comparing,macneil2023experiences,leinonen2023using, bird2022taking, kazemitabaar2023studying, prather2023s}. However, researchers have raised concerns that students may become overly reliant on AI-generated solutions, develop only a superficial understanding of programming concepts, or acquire an inflated perception of their own competence \cite{becker2023programming, finnieansley2022robotscoming, leinonen2023comparing, li2022competition, prather2018metacognitive, prather2023s, denny2024prompt, vaithilingam2022expectation}. Recent studies further suggest that struggling students may be particularly vulnerable to these effects, often relying on GenAI tools in ways that bypass deeper learning processes \cite{margulieux2024srl, prather2024widening}. These concerns are especially relevant in algorithmic problem solving, where conceptual understanding, abstraction, and design reasoning play a central role.

Despite the growing literature on GenAI in computing education, much of the existing work focuses on evaluating the capabilities of LLMs themselves rather than measuring their impact on student learning. A recent large-scale review by Prather et al. found that the majority of published studies investigate code generation performance, AI-assisted programming, or student perceptions, while comparatively little work provides controlled empirical evidence regarding learning outcomes \cite{prather2023robotshere}. Similarly, a recent systematic review by Liu et al.~\cite{liu2025teaching} in algorithms education notes that rigorous controlled studies remain relatively rare in computing education research overall; the authors argue that many open questions remain regarding how algorithms should be taught effectively and equitably, particularly in contexts that require deeper conceptual understanding and problem-solving skills. 

At the same time, the role of students in programming tasks may itself be changing. As GenAI systems become more capable of producing complete solutions, the educational emphasis may increasingly shift from code production toward code comprehension, evaluation, and critique \cite{denny2024explaining, nam2024using, prather2023robotshere}. Recent work argues that understanding, validating, and reasoning about AI-generated code are becoming essential skills for modern programmers \cite{qiao2026systematic}. This shift is particularly important given that modern LLM-based tools, which are freely and openly accessible to students, are already widely used for assignment assistance, and can often generate solutions to introductory programming and algorithms problems with minimal prompting \cite{bird2022taking, finnieansley2022robotscoming, finnie2023my, reeves2023evaluating, savelka2023thrilled}. Compounding this challenge, prior work has noted that AI-generated solutions are difficult to reliably detect using existing plagiarism or AI-detection tools, whose effectiveness remains limited and prone to false positives \cite{orenstrakh2024detecting}. 
Consequently, if the goal is to support genuine learning rather than merely restrict tool use, many researchers argue that simply prohibiting AI usage may be neither practical nor pedagogically effective, and that computing education instead needs instructional approaches that explicitly incorporate and critically engage with AI-generated content \cite{bellettini2023davinci, denny2023conversing, poldrack2023ai, savelka2023large, prather2023robotshere}.

This raises an important pedagogical question: \emph{rather than asking students only to produce solutions themselves, can structured interaction with AI-generated solutions improve conceptual understanding and learning outcomes?}

We investigate the impact of a course-integrated GenAI activity in a junior-level algorithms course. Across a sequence of homework assignments, students participated in one of two instructional conditions: they either (a) solved a challenging algorithmic problem in the traditional manner, or (b) generated, evaluated, and graded an LLM-produced solution to the same type of problem. To study the effects of these activities, we employed a randomized A/B crossover design at the homework working-group level, allowing all groups to experience both instructional conditions at different stages of the semester through a mid-semester treatment switch.

We evaluate the impact of this intervention using several complementary outcome measures, including overall performance on summative assessments, performance on exam questions specifically designed to test conceptual insights emphasized in the GenAI-based assignments, and students’ self-reported perceptions and study strategy adaptations throughout the course.

Our study contributes empirical evidence to the growing discussion surrounding GenAI in computing education. Rather than evaluating the raw capabilities of LLMs, we examine how a specific AI-mediated pedagogical intervention influences student learning outcomes in an algorithms context. More broadly, our findings advance ongoing efforts to rethink assessment and instructional design in the era of Generative AI.

\section{Related Work}

\subsection{Generative AI and the Changing Validity of Traditional Computing Assessments}

Community syntheses argue that the rapid improvement of LLMs challenges the validity of most out-of-class assessments, since high quality solutions and explanations can be generated with minimal effort \cite{prather2023robotshere,prather2025beyond}. Empirical studies have documented this over a large number of languages and assessment formats. For example, LLMs are capable of solving a large portion of introductory programming tasks (and, in some cases, entire programming course assessment suites) \cite{savelka2023thrilled,finnieansley2022robotscoming}. Evaluations in undergraduate programming courses in different languages report that LLMs can generate plausible solutions quickly, motivating a redesign of assessments, learning tasks, and policy \cite{ouh2023chatgpt}. In domains that require structured reasoning, LLMs also show increasing competence. For example, Poulsen et al.~\cite{poulsen2024solving} showed that newer models perform substantially better on proof-related tasks, although important limitations persist. Collectively, these studies suggest that assessment redesign is a pedagogical necessity, rather than merely an academic-integrity response.

\subsection{Pedagogical Responses: Verification, Critique, and Responsible GenAI Use}

A growing number of classroom interventions have responded by foregrounding skills that are more durable under ubiquitous code generation: understanding, testing, debugging, decomposition, and critique \cite{vadaparty2024cs1llm,chandrashekar2026demystify}. This emphasis on structured integration also appears in course-level frameworks that introduce GenAI through explicit norms, guided activities, and reflection on appropriate use, rather than treating LLMs as an unsupervised add-on to existing programming coursework \cite{dickey2024ailab}. Similarly, tool-based interventions attempt to cultivate skepticism by confronting students with mixed- or low-quality AI suggestions and requiring them to identify errors (thus ``operationalizing'' critical use as an assessable learning objective) \cite{macneil2025fostering}. Complementary observational studies, e.g., Smith et al.~\cite{smith2025spotting} and Prather et al.~\cite{prather2024widening}, highlight why scaffolding matters: when students encounter LLM errors, they vary widely in their ability to detect missteps, calibrate trust, and decide on their next actions. 
Together, these findings motivate tasks in which evaluation is part of the design: the learning objective is not merely to produce an answer, but to develop justified confidence through verification and explanation \cite{prather2025beyond, tankelevitch2024metacognitive}.

\subsection{Metacognition and Self-Regulated Learning in GenAI Evaluation Tasks}

Metacognition and self-regulated learning provide a principled lens for why evaluating a GenAI solution might support learning: it can require learners to monitor understanding, detect gaps, and regulate strategy (e.g., rederiving, testing, and seeking counterexamples) \cite{loksa2022metacognition}. Formative feedback theory argues that learners benefit when feedback processes help them generate internal feedback and act on it, rather than passively receiving correctness signals \cite{nicol2006formative}. LLM-mediated student workflows increase these demands: human-AI interaction research characterizes GenAI use as requiring planning, evaluation, trust calibration, and strategy control \cite{tankelevitch2024metacognitive}. Empirical computer science education studies, e.g., Prather et al.~\cite{prather2024widening} and Margulieux et al.~\cite{margulieux2024srl}, suggest these regulatory skills mediate whether GenAI helps or harms: some learners accelerate their learning by using GenAI strategically while others struggle, accept incorrect suggestions, or experience an ``illusion of confidence.'' Altogether, this evidence motivates interventions that \textit{explicitly} scaffold monitoring and evaluation (e.g., justifying why an answer is correct, asking for tests or invariants, or requiring reflection on when the AI was reliable) rather than assuming these behaviors will emerge spontaneously \cite{chandrashekar2026demystify,macneil2025fostering}.

\subsection{Learning from Solutions: Worked Examples, Erroneous Examples, and Self-Explanation}

Evaluation-centered GenAI tasks can be grounded in the worked-example literature. A review by Muldner et al.~\cite{muldner2022workedexamples} focused on programming activities concludes that worked examples can support learning, but outcomes strongly depend on how students engage with the example (tracing/predicting/explaining/debugging), and on alignment to learners' prior knowledge. 
This connection is reinforced by self-explanation research, which shows that learners who generate \textit{principled} explanations while studying a worked solution develop more transferable understanding \cite{chi1989selfexplanations}. Experimental work in learning science, e.g., Durkin and Rittle-Johnson~\cite{durkin2012effectiveness} and Booth et al.~\cite{booth2013using}, has found that studying incorrect examples (usually in contrast with correct ones) can improve conceptual understanding, especially when paired with explanation prompts that direct attention to critical features. These results provide a more theoretical basis for pairing an LLM-solution evaluation with structured prompts (e.g., ``state the claim, identify first invalid step, justify why it fails, and propose a corrected invariant/argument'') rather than treating ``spot the bug'' as a purely unguided activity \cite{chi1989selfexplanations,booth2013using}.

\subsection{Learning by Reviewing: Peer Assessment and Pedagogical Code Review}

Evaluating others' work is a well-established pedagogical structure in computing education, most prominently through peer assessment and pedagogical code review. A systematic review of peer code review in higher education reported common benefits (exposure to alternative solutions, increased attention to quality criteria) and recurring difficulties such as novice reviewers' limited expertise, reliability concerns, and the need for rubrics and scaffolding \cite{indriasari2020peercodereview}. Empirical studies of Pedagogical Code Review (PCR) environments by Hundhausen et al.~\cite{hundhausen2010design,hundhausen2013talking} found that the design of the review activity itself and the review interface shaped whether or not students engaged in substantive critique. Related work by Reily et al.~\cite{reily2009twopeers} also suggests that aggregating multiple peer reviews can improve accuracy, and that participating in the review process can yield learning benefits for reviewers themselves, though outcomes depend on implementation details.
For GenAI evaluation tasks, this body of work suggests that grading an AI solution should not be framed as a trivial addition; it is a form of review activity that requires (1) explicit criteria, (2) accountability for students' justification, and (3) carefully calibrated assignment difficulty so novices can succeed at meaningful diagnosis.

\subsection{LLM-Generated Instructional Artifacts and Reproducibility Challenges}

LLMs are increasingly studied not only as solution generators, but also as feedback providers and evaluators. Systematic evidence on automated grading and feedback tools highlights both the potential value of scalable feedback and persistent concerns about reproducibility and evaluation datasets \cite{messer2024automated}. Recent work by Smith and Zilles~\cite{smith2024code} has explored LLM-mediated grading of free response explanations (e.g., evaluating ``explain in plain English'' answers) and reported moderate agreement patterns with human graders and systematic disagreement patterns that are important for instructional validity. Other classroom tools use LLMs to generate learning content (such as worked examples) and evaluated those experiences via expert review and student studies, e.g., Jury et al.~\cite{jury2024evaluating}. Related platform work by Sinha et al.~\cite{sinha2024boilertai} has also explored instructor-facing uses of GenAI, such as forum support systems in which model-generated responses are reviewed and revised before reaching students. This highlights a complementary design space in which AI is mediated through instructional workflows rather than used directly by students without oversight. However, all of these approaches struggle with LLM variability: outputs can differ with prompts, context, and model version. These differences can threaten fairness and replicability if students receive (meaningfully) different artifacts to evaluate \cite{prather2025beyond,macneil2025fostering}. Practical interventions therefore increasingly recommend explicit controls over factors such as prompts, ``chat freshness,'' artifact reporting, and instructor-generated canonical outputs, as well as explicit instruction on verification approaches \cite{chandrashekar2026demystify,macneil2025fostering}.

\subsection{Gaps in Existing Evidence for Post-Introductory Computing Courses}

Recent synthesis studies, e.g., Bouvier et al.~\cite{bouvier2026rest} and Prather et al.~\cite{prather2025beyond}, explicitly call out the relative scarcity of evidence in post-introductory courses where abstraction and proof-like reasoning play a bigger role. In these settings, the question is usually not ``Can the students write code?'' but rather, ``Can the students justify correctness, reason about tradeoffs, and transfer abstract paradigms?'' Evidence that LLMs can increasingly handle structured reasoning tasks (with limitations) strengthens the case that upper-division courses must develop evaluation-centered learning objectives rather than relying on traditional creation-centric tasks alone \cite{poulsen2024solving,prather2023robotshere}. Our study responds to this gap by examining an evaluation-centered intervention in a junior-level algorithms course using a randomized crossover design and outcomes tied to summative exam performance and structurally aligned reasoning tasks.
\section{Methods}

This study employed a randomized A/B crossover experimental design to examine the relationship between grading AI-generated solutions and students' ability to internalize difficult, abstract concepts in a junior-level ``Introduction to the Analysis of Algorithms'' course. Specifically, for each homework assignment, approximately half of the students were asked to solve a problem independently, while the remaining students were tasked with evaluating a GenAI-generated solution to the same problem. Student outcomes under these two instructional approaches were then compared using performance measures from the midterm and final examinations, as well as students' self-reported perceptions of the two types of exercises.

The course covers classical algorithm design paradigms, including divide-and-conquer, dynamic programming, greedy algorithms, maximum flow, and reductions, and is traditionally most challenging during the post-midterm portion, when students are required to engage in increasingly complex abstract reasoning. The study was conducted during the Spring 2025 semester at Purdue University.

\subsection{Research Questions}

This study investigates how engaging with AI-generated solutions as a grading task, compared to traditional problem solving, relates to students' learning outcomes and perceptions in a junior-level algorithms course. Specifically, we address the following research questions:

\begin{question}
    How does evaluating GenAI-generated solutions, relative to directly solving problems, relate to students' performance on summative assessments (midterm and final examinations)?
\end{question}

\begin{question}
    To what extent does participation in GenAI-based grading activities relate to students' ability to internalize and transfer abstract algorithmic reasoning, as reflected in performance on exam problems structurally aligned with prior homework exercises?
\end{question}

\begin{question}
    How do students perceive the helpfulness of grading GenAI-generated solutions for learning algorithmic concepts, compared to completing problems independently?
\end{question}

\begin{question}
    Does participation in GenAI-based grading activities relate to self-reported changes in students' study behaviors and preparation strategies for examinations?
\end{question}

\subsection{Participants and Group Formation}

A total of 227 undergraduate students were enrolled in the course. After excluding seven students who audited the course, received incompletes, or did not complete the final examination, the final analytic sample consisted of 220 students.

At the beginning of the semester, students were informed that they would be divided into two groups, each receiving a slightly different set of homework assignments. To support existing collaboration practices and reduce cross-group contamination, students were asked to self-report their regular working groups (up to six peers), with the assurance that members of the same working group would be assigned to the same set of homework questions. This approach was intended to ensure that students who routinely collaborated remained together, thereby minimizing interactions across conditions (e.g., in cases where one group was tasked with solving a problem while the other evaluated AI-generated solutions for the same problem).
While collaboration within reported working groups was permitted under the course collaboration policy, students submitted homework individually and were expected to write up their own solutions. 

Once working groups were identified, they were randomly assigned to one of the two experimental conditions. The resulting groups were labeled \emph{Monte Carlo} and \emph{Las Vegas}, referencing two classes of randomized algorithms. These labels were intentionally chosen to avoid any implied ordinal or performance-based hierarchy between conditions, thereby reducing the risk of expectancy effects or motivational bias.

\subsection{Ethics and Consent}

The study protocol was reviewed and approved by the Purdue University Institutional Review Board (IRB-2025-39, approved January 21, 2025). Survey participation was voluntary, and survey responses were collected anonymously. Course performance data were analyzed after de-identification.

\subsection{Experimental Design}

We employed an A/B crossover design to compare learning outcomes associated with solving problems versus evaluating GenAI-generated solutions. The course included six homework assignments, one midterm examination, and one final examination. Homework assignments were released biweekly, each with a two-week submission window, and each focused on a distinct course topic: (1) mathematical induction, (2) divide-and-conquer algorithms, (3) dynamic programming, (4) graph algorithms, (5) max-flow/min-cut, and (6) reductions.

Each homework assignment consisted of four exercises of increasing difficulty. The first three exercises were traditional problem-solving tasks in which students were required to produce complete solutions; these exercises were identical across both experimental conditions and were each worth 30 points. The fourth exercise was deliberately designed to be substantially more challenging, such that contemporary GenAI systems (e.g., ChatGPT) could not reliably produce correct solutions: before release, we tested each prompt in multiple independent ChatGPT-4o sessions and selected problems for which generated solutions were consistently incomplete or incorrect. This exercise was worth 10 points for students in both conditions. The number of points was chosen so that no group gained a significant advantage over the other.

During the first half of the semester (Homework 1--3), students assigned to the \emph{Monte Carlo} condition were required to solve the fourth exercise directly. This typically involved designing an algorithm, formally proving its correctness, and analyzing its runtime. In contrast, students in the \emph{Las Vegas} condition were provided with a standardized prompt to submit to ChatGPT and were then tasked with evaluating the generated solution, including identifying and explaining any errors or omissions.

Below, we present an example of a homework problem as assigned to the \emph{Monte Carlo} group, along with the corresponding instructions provided to the \emph{Las Vegas} group. For clarity in the analyses that follow, the Las Vegas condition is reported as Group A and the Monte Carlo condition as Group B.

\begin{assignment}
    \textbf{Monte Carlo Assignment. }You are trying to build an amusement park ride out of spare parts that you found at the University Surplus Store. Each part has a certain shape that we represent as a sequence of pluses and minuses, as in: ``$++-----+$,'' representing a part of length 8, that climbs up 2 feet, drops down 5 feet, then climbs 1 foot (ending 2 feet below where it started). Formally, you are given as input an array of $n$ strings. Each string is of length at most $\ell$, and it consists only of $\{+,-\}$ symbols. You want to find the length of the longest ride you can make using (a subset of) these parts, where a ride must \emph{start} on the ground, must \emph{end} on the ground, and must \emph{never go below ground}. For example, if you are given $[--+++, --+++, ++,  +, ---]$, then the longest ride you can make is of length 10 and starts with the 3rd part, then the 2nd part, then the 5th part, ``$++--+++---$.'' Devise a polynomial time algorithm, which given as input the array of parts, returns the length of the longest ride. Prove the correctness and runtime of your algorithm using clear and concise arguments.
\end{assignment}

\begin{assignment}
    \textbf{Las Vegas Assignment. }Create a new chat in ChatGPT-4o and ask it to provide a (succinct) solution to the following problem. Provide a link to your chat (using the ``Share Chat'' functionality) and a (legible) screenshot of the response. You can include this in \LaTeX{} using the figure environment. If you are working in a study group, everyone in the group can submit the same screenshot.

        ``You are trying to build an amusement park...''
        
 \noindent Grade ChatGPT's solution from Part (a). Specifically, if both the algorithm and its proof of correctness work, state that this is the case. If there is a bug in the proof, point to the bug, and argue whether this is a minor bug (e.g., a calculation error) but the proof is more or less correct, or whether it is a major bug (i.e., a crucial step is wrong or missing).
\end{assignment}

In the GenAI-evaluation condition, ``grading'' did not mean only assigning a numerical score. Students were expected to evaluate the correctness of the generated algorithm and proof, identify any substantive errors or omissions, and explain why those errors affected the validity of the solution. In particular, students were asked to distinguish minor issues, such as arithmetic or presentation errors, from major issues, such as an incorrect algorithm, an invalid recurrence, a missing invariant, or a flawed proof step. Students were not required to provide a complete corrected solution, although they could do so as part of their explanation.

To reduce variation across GenAI-generated artifacts, students in the evaluation condition were instructed to begin a new ChatGPT-4o chat and use the provided prompt without modifying the problem statement. Students submitted both a share link and a screenshot of the generated response, allowing the instructional staff to verify that a GenAI-generated solution had been produced and evaluated. We did not require all students to evaluate an identical instructor-generated output; instead, the intervention intentionally reflected a realistic setting in which students interact directly with GenAI tools. Despite variation across individual responses, generated solutions for each assignment tended to fall into a small number of recurring error patterns, summarized in \Cref{tab:summary_grades}. Students were instructed to evaluate the initial generated solution rather than iteratively prompting the model until it produced a correct answer.

The exercise described above requires first preprocessing the input by sorting the list of parts according to the minimum height each part can reach independently. The solution then proceeds by constructing a two-dimensional dynamic programming table, where the first dimension corresponds to the part under consideration and the second dimension represents the height of the roller coaster.
Across all homework assignments, we observed that the GenAI-generated solutions consistently fell into two to three distinct categories, each characterized by recurring and systematic errors. In the present exercise, the GenAI solution failed to sort the input and, in many instances, incorrectly formulated the dynamic programming approach by using a one-dimensional table instead of the required two-dimensional structure.
After the first three homework assignments were completed, students took the midterm examination. The midterm was administered in person, proctored by the course instructor and teaching assistants, with closed books and the allowance of a single handwritten notes sheet per student. The exam assessed material covered in the first three assignments. Notably, one midterm problem required reasoning closely aligned with that of a prior homework exercise, specifically, an exercise for which one of the student groups had evaluated a GenAI-generated solution. The corresponding midterm problem is presented below.

\begin{assignment}
    \textbf{Midterm Exercise. } You have decided to take a job trading in Wall Street. Using your superpower, algorithm design, you figure out $n$ trading strategies, where strategy $i$, when executed, gives you a profit of $v_i \geq 0$ million dollars.  You want to execute your strategies over $n$ days. Strategies cannot be re-used, so once you've used a strategy $i$, you cannot use it again. 

    You would really like to use all your strategies, but you are scared that you'll attract too much attention to yourself, and then others will start stealing your strategies, taking your competitive advantage away. Therefore, you decide that on the $k$-th day, you should make sure to have accumulated a profit of at most $5 k$ million dollars so far. That is, the strategy on the first day you should make at most $5$ million dollars, on the first two days you should make at most $10$ million dollars, and so on. To make sure this constraint is satisfied, some days you show up at the office, and execute a strategy that does nothing. Unfortunately, your boss notices when you do this, so, in order to keep your job, you want to maximize the number of strategies (``real'' strategies, from the set $\{ s_1, \dots, s_n\}$) you execute, while keeping a low profile, and, of course, making a lot of money.

    Devise an efficient algorithm that, given as input $n$ strategies $s_1, \dots, s_n$ with profits $v_1, \dots, v_n$ (in millions of dollars), constructs a play-book for $n$ days which describes whether on day $i$ you execute some strategy $s_j$ (at most one strategy) or do nothing, such that (1) you maximize the number of strategies from the set $s_1, \dots, s_n$, (2) for all $k = 1, \dots, n$ the profit you accumulate in the first $k$ days is at most $5k$ million dollars, (3) the total profit you make is exactly $x$ million dollars, for some $x < 5n$.
Prove the correctness of your algorithm, and analyze its runtime. 
\end{assignment}

For the exercise described above, the solution requires first sorting the input based on the profit associated with each strategy, followed by the construction of a three-dimensional dynamic programming table. The first dimension corresponds to the day under consideration, the second to the strategy being evaluated, and the third to the cumulative profit achieved. This line of reasoning closely mirrors the structure and logic of the roller coaster exercise introduced earlier in the homework assignment.

During the second half of the semester, the roles of the two groups were reversed. Students in the Las Vegas condition were tasked with directly solving the exercise, whereas students in the Monte Carlo condition evaluated a GenAI-generated solution. At the end of the semester, students completed a final examination covering all course topics, with a stronger emphasis on material from Homework Assignments 4--6. As in the midterm, one exam problem was intentionally designed to closely align with the structure and reasoning of a prior homework exercise, specifically, an exercise for which the Monte Carlo group had evaluated a GenAI-generated solution.

\subsection{Data Collection and Instruments}\label{subsec:data collection}

We analyzed multiple sources of quantitative and self-reported data: (1) item-level and aggregate scores from all homework assignments and the two course examinations; (2) final course percentage grades; (3) responses from a pre-midterm survey assessing topic-specific comfort and confidence ($N=208$); and (4) responses from a post-final survey examining students' perceived helpfulness of GenAI-based grading activities and self-reported changes in study habits ($N=200$).

All homework assignments and examinations were graded by the course instructor and the teaching assistant team, which consisted of seven graduate teaching assistants (GTAs) and seventeen undergraduate teaching assistants (UTAs). All UTAs had previously completed the course and earned a final grade of A or A+. UTAs together with GTAs were responsible for grading the first three (lower-difficulty) exercises on each homework assignment, whereas the fourth, most challenging exercise on each assignment was graded exclusively by the GTAs. Prior to grading, all graders responsible for a given exercise met to jointly develop a grading rubric and align on evaluation criteria, with the goal of minimizing inter-grader variability. All homework and exam grading was conducted using the Gradescope platform. For the examinations, graders additionally worked in a shared physical space and graded concurrently to further reduce grading inconsistencies.

The grading criteria for the fourth exercise differed by condition but emphasized the same underlying algorithmic insight. For students solving the problem directly, credit was awarded for a correct algorithm, proof of correctness, and runtime analysis. For students evaluating a GenAI-generated solution, credit was awarded for accurately diagnosing the generated solution, identifying major correctness issues, and justifying the evaluation with reference to the algorithm, proof, or runtime analysis. Thus, the two conditions were not graded using identical rubrics, but both were designed to assess engagement with the same core algorithmic idea.

Survey data were collected using a university-supported platform that required students to authenticate with their institutional credentials. The surveys did not collect identifying information, ensuring anonymity and encouraging honest responses. Approximately one hour prior to the midterm examination, at a point when nearly all exam preparation had been completed, we administered a Qualtrics survey to assess students' self-reported comfort and confidence across the core algorithmic topics covered on the midterm. Response rates were high (Group A: $n=101$; Group B: $n=107$; total $N=208$, 95\% response rate). All items used a five-point Likert scale (1 = Very uncomfortable / Not confident; 5 = Very comfortable / Very confident). Due to technical constraints, an equivalent survey could not be administered prior to the final examination.

Immediately following the final examination, we administered a second Qualtrics survey to both experimental groups (Group A: $n=97$; Group B: $n=103$; total $N=200$, 91\% response rate). The survey included two closed-ended items. The first assessed students' perceived helpfulness of grading a GenAI-generated solution relative to completing the problem themselves, measured on a five-point Likert scale (1 = Very unhelpful; 5 = Very helpful). The second item examined whether the modified homework assignments influenced students' exam preparation strategies, using a three-point nominal scale (1 = No; 2 = Somewhat; 3 = Yes).

\subsection{Statistical Analysis}

For each outcome measure, we followed a three-step analytic workflow:
\begin{enumerate}
    \item assumption screening using the Shapiro--Wilk test for normality and Levene's test for homogeneity of variance;
    \item test selection based on the results of assumption checks and the measurement scale of the data; and
    \item effect size reporting to assess practical significance.
\end{enumerate}

For normally distributed, interval-scale data with homogeneous variances, we employed independent-samples $t$-tests and reported Cohen's $d$. When these assumptions were violated, or when outcomes were ordinal, we used appropriate nonparametric alternatives, including the Mann--Whitney $U$ test (two independent groups), the Wilcoxon signed-rank test (paired samples), and the Kruskal--Wallis $H$ or Friedman tests (three or more related samples). Corresponding effect sizes were reported using rank-biserial $r$ or Kendall's $W$, as appropriate. Two-sample Kolmogorov--Smirnov tests were additionally used as shape-based robustness checks.
Associations between ordinal variables were analyzed using Spearman's $\rho$, while categorical frequency data were analyzed using $\chi^{2}$ tests with Cram\'er's $V$ reported as the effect size. For the problem-level analysis of final examination performance, where score distributions exhibited substantial skew, we report the median and interquartile range (IQR) and rely on nonparametric tests.

To control family-wise error rates in post hoc comparisons, we applied Holm's sequential Bonferroni correction. All statistical tests were two-tailed with a significance threshold of $\alpha = .05$.

\section{Results}

\begin{table*}
\makebox[\columnwidth][c]{

    \begin{tabular}{|c | c | c | c | c | c |} 
     \hline \rowcolor{Gainsboro!80}
     Topic & \shortstack{Key Solution Idea} & \shortstack{LLM's Mistake} & \shortstack{GenAI Group\\Score Summary\\(group, $\mu$, median, $\sigma$)} & \shortstack{Non-GenAI Group \\ Score Summary \\ (group, $\mu$, median, $\sigma$)} & \shortstack{Total \\ Score Summary \\ ($\mu$, median, $\sigma$)} \\
     \hline\hline
     Induction & \shortstack{Add-Subtract value and \\ use triangle inequality \\ to produce bound} & \shortstack{Wrong proof\\of correctness} & A, 5.41, 6.00, 3.89 & B, 6.23, 6.80, 2.77 & 5.82, 6.00, 3.38\\ 
     \hline
     \shortstack{Divide \&\\Conquer} & \begin{tabular}{c} Even \& odds cannot \\ create invalid pairs, \\adding \& dividing does not \\ create new invalid pairs \end{tabular} & Wrong Algorithm & A, 6.38, 8.00, 3.46 & B, 5.95, 6.00, 3.28 & 6.16, 7.11, 3.37\\
     \hline
     \shortstack{Dynamic\\Programming} & \shortstack{Sort entries by an auxiliary \\ value, create 2D DP table} & Wrong Algorithm & A, 6.11, 7.00, 3.59 & B, 5.36, 5.60, 3.33 & 5.73, 6.00, 3.47\\
     \hline \rowcolor{Gainsboro!40}
     Average & -  & - & A, 5.97, 6.33, 2.74 & B, 5.85, 6.00, 2.45 & 5.91, 6.00, 2.59 \\
     \hline \hline

     \shortstack{Graph\\Problems} & \begin{tabular}{c} Create multiple copies \\ of given graph and \\
     connect copies through \\ appropriate edges \end{tabular} & Wrong Algorithm & B, 6.41, 7.50, 3.51 & A, 2.99, 1.50, 3.26 & 4.71, 4.00, 3.79\\
     \hline
     \shortstack{Network\\Flow} & \begin{tabular}{c} Problem is equivalent \\ to finding a partition \\ of nodes to disjoint cycles \\ in a bipartite graph \end{tabular} & \shortstack{Wrong proof\\of correctness} & B, 7.21, 8.00, 3.38 & A, 3.90, 3.59, 3.07 & 5.57, 6.00, 3.62\\ 
     \hline
     Reductions &\begin{tabular}{c} We can enforce equality \\ on first inequality by \\choosing tight number \\for the second inequality \end{tabular} & Wrong Algorithm & B, 6.54, 8.00, 3.85 & A, 5.55, 6.00, 3.31 & 6.05, 7.00, 3.62\\ 
     \hline \rowcolor{Gainsboro!40}
     Average  & -  & - & B, 6.72, 7.33, 2.67 & A, 4.15, 3.68, 2.56 & 5.44, 5.63, 2.91 \\
     \hline \hline

     \shortstack{Midterm\\Question} & \begin{tabular}{c} Sort entries by an auxiliary \\ value, create 3D DP table \end{tabular}  & - & A, 1.62, 0.00, 3.29 & B, 1.95, 0.00, 3.35  & 1.79, 0.00, 3.32\\
     \hline \rowcolor{Gainsboro!40}
     \shortstack{Midterm\\Overall} & -  & - & A, 60.69, 62.00, 14.45 & B, 61.78, 62.00, 13.41 & 61.24, 62.00, 13.91 \\
     \hline
     \shortstack{Final\\Question} & \begin{tabular}{c} Create multiple copies of\\given graph and connect\\copies through appropriate edges \end{tabular}  & - & B, 5.43, 2.00, 5.22  & A, 5.79, 6.00, 5.15 & 5.61, 6.00, 5.17 \\
     \hline \rowcolor{Gainsboro!40}
     Final Overall & -  & - & B, 59.06, 56.50, 16.72  & A, 58.18, 59.00, 17.07 & 58.62, 58.00, 16.86\\
     \hline
    \end{tabular}
    }
    \caption{HW grades are out of 10, Midterm Question Grades are out of 15, Midterm Overall Grades are out of 101, Final Question Grades are out of 15, Final Overall Grades are out of 113. 
    ``GenAI Group'' denotes the group that evaluated a GenAI-generated solution for the corresponding homework topic or prior homework block; ``Non-GenAI Group'' denotes the group that solved the corresponding problem directly.
    The selected midterm and final questions required insight from one of the GenAI-evaluation questions during the prior 3 homeworks.
    }
    \label{tab:summary_grades}
\end{table*}

A summary of homework LLM questions and midterm and final questions and grades can be found in \Cref{tab:summary_grades}.

\subsection{Baseline Equivalence and Internal Validity}

To assess whether the two groups were comparable prior to the GenAI-grading intervention, we examined a baseline achievement measure derived from the sum of the three non-AI problems on Homeworks 1--3, rescaled to a 0–10 scale (Group A: $n=109$; Group B: $n=111$). Descriptive statistics indicated similar performance across groups (Group A: $M = 7.91$, $\mathrm{SD} = 1.36$, median = 8.04; Group B: $M = 8.04$, $\mathrm{SD} = 1.35$, median = 8.30), with overlapping 95\% confidence intervals for the mean (Group A: $[7.65, 8.17]$; Group B: $[7.79, 8.30]$). 

Assumption screening revealed departures from normality for both groups (Shapiro--Wilk: $W_{\mathrm{A}} = 0.94$, $W_{\mathrm{B}} = 0.91$, $p < .001$), while homogeneity of variance was satisfied (Levene's test: $F = 0.15$, $p = .69$). Given the large and balanced sample sizes, we report both parametric and nonparametric comparisons. An independent-samples $t$-test indicated no statistically significant difference between groups, $t(218) = -0.72$, $p = .47$. Consistent with this result, a Mann--Whitney $U$ test likewise showed no significant difference ($U = 5658$, $p = .407$), with a trivial effect size ($r = 0.06$). A two-sample Kolmogorov--Smirnov test further confirmed that the distributions did not differ in shape ($D = 0.097$, $p = .62$).

Taken together, these analyses converge to indicate that the two cohorts were statistically indistinguishable at baseline. This supports the internal validity of subsequent comparisons, suggesting that later differences in outcomes are unlikely to be attributable to pre-existing achievement differences between groups.

\subsection{Primary Outcome: Overall Exam and Course Performance}

With baseline equivalence established, we next examined whether the order in which students experienced the two instructional conditions was associated with differences in overall course performance, as measured by midterm examination scores, final examination scores, overall course percentages, and relative change in exam performance over the semester. As detailed below, performance on all measures remained statistically indistinguishable across groups.

For each high-stakes assessment, we first evaluated assumptions relevant to test selection. Shapiro--Wilk tests failed to reject normality for midterm scores, final scores, normalized change scores (Final -- Midterm), and overall course percentages in both groups (most $p > .35$, one $p > .07$). Levene's tests further indicated homogeneity of variances for all outcomes (all $p > .19$). Given that these assumptions were satisfied, independent-samples $t$-tests were used as the primary inferential procedure. For all comparisons, we report 95\% confidence intervals (CIs) for the mean differences and Cohen's $d$ as a measure of effect size.

Across all summative measures, group means were closely aligned. On the midterm examination (maximum score = 101), Group A achieved a mean score of $M = 60.7$ ($\mathrm{SD} = 14.5$), while Group B achieved a mean score of $M = 61.8$ ($\mathrm{SD} = 13.4$). On the final examination (maximum score = 113), Group A obtained a mean of $M = 58.2$ ($\mathrm{SD} = 17.1$), compared to $M = 59.1$ ($\mathrm{SD} = 16.7$) for Group B. Final course percentages were similarly comparable, with Group A averaging $M = 63.9$ ($\mathrm{SD} = 12.1$) and Group B averaging $M = 65.4$ ($\mathrm{SD} = 11.4$).

To capture relative changes in exam performance over time, we additionally computed a normalized change score (Final -- Midterm, scaled to 10 points). Group A exhibited a mean change of $M = -0.86$ ($\mathrm{SD} = 1.3$), while Group B exhibited a mean change of $M = -0.89$ ($\mathrm{SD} = 1.1$). The negative values reflect the differing exam maxima and the greater difficulty of post-midterm material rather than an absolute decline in learning.

Inferential analyses revealed no statistically significant differences between groups on any summative performance measure. Midterm examination scores did not differ significantly between groups, $t(218) = -0.58$, $p = .56$, with a 95\% CI of $[-4.8, 2.6]$ points and a negligible effect size (Cohen's $d = -0.08$). Final examination scores likewise showed no significant difference, $t(218) = -0.39$, $p = .70$, 95\% CI $[-5.4, 3.6]$ points, $d = -0.05$.

The normalized change in exam performance (Final -- Midterm) was also statistically indistinguishable across groups, $t(218) = 0.19$, $p = .85$, 95\% CI $[-0.3, 0.3]$ points, $d = 0.03$. Similarly, overall course percentages did not differ significantly, $t(218) = -0.95$, $p = .34$, 95\% CI $[-4.6, 1.6]$ points, with a negligible effect size ($d = 0.13$).

Finally, distributions of letter grades were comparable across groups, $\chi^{2}(9) = 9.69$, $p = .38$, with a small effect size (Cram\'er's $V = .07$), further supporting the absence of meaningful differences in overall course outcomes.

\subsection{Performance in GenAI-Aligned Exam Question}

In each examination (midterm and final), we included a problem whose solution required the same core techniques and sequence of logical steps as a prior homework exercise. On the midterm aligned question, Group A had previously evaluated a GenAI-generated solution to the corresponding homework problem, while Group B had solved the problem directly. On the final aligned question, this mapping was reversed. This crossover structure therefore provides two within-group observations, enabling us to examine whether prior participation in GenAI-based grading is associated with differences in the internalization and transfer of algorithmic reasoning.

\subsubsection{Analytic approach and descriptive statistics.}

Shapiro--Wilk tests indicated that all four samples deviated from normality ($W=.57$--$.87$, $p<10^{-7}$), while Levene's tests supported the assumption of homogeneous variances (midterm $p=.46$, final $p=.77$). Accordingly, group comparisons were conducted using the Mann--Whitney $U$ test as the primary inferential procedure, with two-sample Kolmogorov--Smirnov (KS) tests used as robustness checks. Effect sizes are reported using rank-biserial $r$, with $\alpha = .05$.
\Cref{tab:summary_grades} summarizes the score distributions for the midterm and final exam items (each out of 15 points). The midterm item exhibited a pronounced floor effect (median = 0 in both groups), whereas the final item showed higher central tendency and greater dispersion, consistent with improved performance over the course of the semester. The midterm item's pronounced floor effect (median = 0 in both groups) limits its sensitivity to detect group differences; the transfer comparison is therefore carried primarily by the final-exam item, and conclusions about midterm transfer should be regarded as inconclusive rather than null.

\subsubsection{Inferential results.}

For the \textbf{midterm}, the Mann--Whitney test indicated no statistically significant difference between groups ($U = 5{,}644.5$, $p = .30$), with a trivial effect size ($r = -0.058$). The two-sample Kolmogorov–Smirnov test yielded a consistent result ($D = .067$, $p = .95$). Similarly, for the \textbf{final}, no statistically significant difference was observed on the final exam item ($U = 6{,}191.5$, $p = .76$), with a negligible effect size ($r = .020$). The KS test again corroborated this result ($D = .073$, $p = .90$). Across both assessment points, we do not observe evidence of differences in performance associated with prior participation in GenAI-based grading activities.

\subsection{Final Exam Performance by Problem Category}\label{sec:final_exam}

To examine whether the timing of GenAI-assisted grading (first- versus second-half assignments) was associated with end-of-course performance, we partitioned the final examination into three categories: first-half, second-half, and hybrid problems. For each student, we computed a percentage score within each category by summing earned points and normalizing by the maximum possible points for that category.

For \textbf{first-half problems}, Group A achieved $M = 71.43\%$ ($\mathrm{SD} = 17.31$; $\tilde{x} = 71.88\%$; $\mathrm{IQR} = 25.00$), while Group B achieved $M = 73.23\%$ ($\mathrm{SD} = 18.28$; $\tilde{x} = 75.00\%$; $\mathrm{IQR} = 31.25$). 

For \textbf{second-half problems}, performance was lower in both groups: Group A obtained $M = 45.93\%$ ($\mathrm{SD} = 17.41$; $\tilde{x} = 45.39\%$; $\mathrm{IQR} = 25.00$), and Group B obtained $M = 46.36\%$ ($\mathrm{SD} = 16.73$; $\tilde{x} = 44.08\%$; $\mathrm{IQR} = 25.66$). 

For \textbf{hybrid problems} (reported descriptively), Group A achieved $M = 51.98\%$ ($\mathrm{SD} = 21.92$; $\tilde{x} = 47.37\%$; $\mathrm{IQR} = 21.05$), and Group B achieved $M = 53.22\%$ ($\mathrm{SD} = 21.85$; $\tilde{x} = 47.37\%$; $\mathrm{IQR} = 26.32$).

\subsubsection{Inferential analyses.}
Shapiro--Wilk tests indicated non-normal distributions for both categories ($p < .05$ for all), while Levene's tests supported the assumption of homogeneous variances (first-half: $p = .384$; second-half: $p = .600$). Accordingly, group comparisons were conducted using Mann--Whitney $U$ tests, with rank-biserial $r$ reported as the effect size.

For first-half problems, the Mann--Whitney test indicated no statistically significant difference between groups ($U = 5{,}624.50$, $p = .368$), with a trivial effect size ($r = -0.061$). The 95\% confidence interval for the median difference was $[-9.38,\, 3.13]$.

For second-half problems, similarly, no statistically significant difference was observed ($U = 5{,}932.00$, $p = .804$), with a negligible effect size ($r = -0.017$). The 95\% confidence interval for the median difference was $[-5.92,\, 6.58]$.

\subsection{Change in Homework Performance} \label{sec:hw_change}

We next examined whether GenAI-based grading was related to changes in students' homework performance over the course of the semester. For each student, we computed two within-student difference scores: (1) the change in performance on the GenAI-graded problem across the two halves of the course (with positive values indicating higher performance during periods of GenAI-based grading), and (2) the corresponding change in performance on the three problems solved independently.

\subsubsection{Descriptive statistics.}

\begin{center}
    \begin{tabular}{lcc}
        Metric & Group A ($M \pm \mathrm{SD}$) & Group B ($M \pm \mathrm{SD}$) \\
        \midrule
        GenAI Questions (\%)  & $18 \pm 25$ & $9 \pm 23$ \\
        Non-GenAI Questions (\%) & $3 \pm 14$ & $-3 \pm 13$
    \end{tabular}
\end{center}

Descriptively, Group A exhibited a larger positive change on the GenAI-graded problems and a modest advantage on the student-solved problems.

\subsubsection{Assumption checks.}
Shapiro--Wilk tests indicated deviations from normality in three of the four distributions ($p < .05$), with the exception of Group A's GenAI-graded difference scores. Levene's tests supported the assumption of homogeneous variances across groups ($p > .11$). Given these mixed normality results, we report nonparametric analyses as the primary inferential procedure, supplemented by parametric tests for the GenAI-graded problems as a robustness check.

\subsubsection{Inferential analyses.}
For the GenAI-graded problems, the Mann--Whitney test indicated a statistically significant difference between groups ($U = 7{,}046$, $p = .035$), with a small effect size ($r = .14$). A confirmatory independent-samples $t$-test yielded a consistent result, $t(218) = 2.90$, $p = .004$, with a small effect size (Cohen's $d = .39$).

Two cautions apply in interpreting this difference. First, the two conditions were not graded with identical rubrics (see~\Cref{subsec:data collection}), so the comparison partly reflects differing scoring structures for solving versus evaluating, not solely a difference in learning. Second, because randomization occurred at the working-group level, this student-level result is the finding most sensitive to within-group dependence (see~\Cref{subsec:threats}).

For the student-solved (non-GenAI) problems, the Mann--Whitney test likewise indicated a statistically significant difference between groups ($U = 7{,}119$, $p = .024$), with a small effect size ($r = .15$). Notably, these tasks were identical across conditions, so this difference cannot be attributed to the intervention; it more likely reflects time-varying factors such as topic difficulty, exam timing, and accumulated course experience across the two halves of the course. This pattern reinforces our broader interpretation that the homework-level differences should not be read as direct evidence of a learning effect from the GenAI activity.

\subsection{Emerging Affective Outcomes (Midterm Self-Efficacy)}

After students had completed three of the six homework assignments under their assigned condition, we administered a Qualtrics survey approximately one hour prior to the midterm examination, at a point when most exam preparation had been completed. The survey achieved a high response rate (Group A: $n = 101$; Group B: $n = 107$; overall $N = 208$, 95\%) and measured students' self-reported \emph{comfort} and \emph{confidence} across the five core algorithmic topics assessed on the midterm: asymptotic analysis, divide \& conquer, dynamic programming, greedy methods, and graph algorithms. Each item used a five-point Likert scale (1 = Very uncomfortable / Not confident; 5 = Very comfortable / Very confident). Due to technical limitations, an equivalent survey could not be administered prior to the final examination.

\subsubsection{Data preparation and analysis.}
Likert-scale responses were recoded to ordinal values (1--5). Internal consistency was assessed using Cronbach's $\alpha$, yielding acceptable reliability for both the Comfort ($\alpha = 0.79$) and Confidence ($\alpha = 0.84$) scales, supporting the use of composite scores. 

All analyses were conducted at $\alpha = .05$ using nonparametric procedures. Between-group differences in Comfort and Confidence composites were evaluated using Mann--Whitney $U$ tests. Within-subject variation across topics was assessed using Friedman tests, followed by Wilcoxon signed-rank post hoc comparisons with Holm correction. Associations between Comfort and Confidence composites were examined using Spearman's rank correlation ($\rho$).

\subsubsection{Results.}
No statistically significant differences were observed between groups on either the Comfort composite ($U = 5{,}788$, $p = .374$) or the Confidence composite ($U = 5{,}152$, $p = .561$) after three assignments under their respective conditions.

Friedman tests indicated significant within-subject variation across topics for both Comfort ($\chi^2(4) = 156.78$, $p < .001$) and Confidence ($\chi^2(4) = 166.50$, $p < .001$). Holm-adjusted post hoc comparisons showed that Asymptotic Analysis (Comfort median = 4; Confidence median = 3--4) and Divide \& Conquer (Comfort median = 4; Confidence median = 3) were rated significantly higher than Dynamic Programming, Greedy Methods, and Graph Algorithms (all medians = 3) on both scales ($p_{\mathrm{holm}} < .001$), with no statistically significant differences among the latter three topics (all $p_{\mathrm{holm}} > .05$).

Finally, a strong positive association was observed between Comfort and Confidence composites (Spearman's $\rho = 0.714$, $p < .001$), indicating substantial concordance between the two measures.

\subsection{End-of-Course Perceptions (Post-Final Survey)}

Immediately after the final exam, we administered a Qualtrics survey to both experimental groups (Group A: $n=97$; Group B: $n=103$; overall $N=200$, 91\% response rate).  Two closed-ended items were included:
\begin{itemize}
  \item \textbf{Helpfulness of Grading.}
    ``How much did grading GenAI's solution (versus completing the problem yourself) help you learn the material?''
    5-point Likert (1 = Very unhelpful $\dots$ 5 = Very helpful).
  \item \textbf{Change in Studying.}
    ``Did the modified homework assignments change how you studied for exams?''
    3-point nominal (1 = No; 2 = Kind of; 3 = Yes).
\end{itemize}
All responses were converted to ordinal codes; group completion rates did not differ, so we proceeded with our preregistered nonparametric tests at $\alpha = .05$.

\subsubsection{Descriptive statistics.}
Both groups reported a neutral median rating on the helpfulness item ($\tilde{x} = 3.0$), indicating that perceptions generally ranged between ``Somewhat unhelpful'' and ``Somewhat helpful.''

With respect to changes in study habits, the majority of students reported no change (154/200, 77.0\%), followed by ``Kind of'' (32/200, 16.0\%) and ``Yes'' (14/200, 7.0\%).

\subsubsection{Between-group comparisons.}
No statistically significant difference was observed between groups in perceived helpfulness of grading (Mann--Whitney $U = 5{,}132$, $p = .728$). 

Similarly, a chi-square test of independence indicated no association between group assignment and reported changes in study habits ($\chi^2(2, N = 200) = 0.97$, $p = .616$), with a small effect size (Cram\'er's $V = 0.07$).

\subsubsection{Association between helpfulness and study-habits change.}
A Kruskal--Wallis test indicated a statistically significant difference in helpfulness ratings across the three study-change categories ($H(2) = 17.74$, $p < .001$). 

Follow-up pairwise Mann--Whitney tests with Bonferroni-adjusted $p$-values  showed that students who reported changing their study habits rated the grading activity as more helpful than those who reported no change (No vs.\ Kind of: $U = 1{,}742$, $p_{\mathrm{adj}} = .019$; No vs.\ Yes: $U = 505$, $p_{\mathrm{adj}} = .002$). The difference between the ``Kind of'' and ``Yes'' categories was not statistically significant ($U = 143$, $p_{\mathrm{adj}} = .122$).

\section{Discussion}

Our study investigated how evaluating AI-generated solutions compares with traditional problem-solving in a junior-level algorithms course. Across several outcome measures, we found no evidence that evaluating GenAI-generated solutions led to higher overall exam performance or course grades than directly solving problems. The two groups were comparable at baseline and remained statistically indistinguishable on midterm scores, final exam scores, course percentages, and changes in exam performance over time. We also found no significant differences on exam questions designed to be structurally aligned with prior homework exercises. Thus, in this implementation, prior exposure to a GenAI-generated solution and its evaluation did not measurably improve transfer to a related exam problem relative to solving the corresponding homework problem directly.

Our findings complicate two common intuitions about GenAI-based coursework. One concern is that asking students to evaluate AI-generated solutions instead of solving problems themselves will substantially weaken learning, because students lose practice constructing complete solutions. A competing hope is that evaluating flawed AI output may strengthen learning by prompting students to inspect, critique, and explain algorithmic reasoning. In our study, neither extreme prediction was supported. Students who engaged in GenAI-solution evaluation did not outperform students who solved the problems directly, but we also found no evidence of broad performance losses on summative assessments or structurally aligned exam questions.

These results suggest that GenAI-evaluation tasks should not be viewed simply as weaker substitutes for problem solving, nor as automatically superior forms of metacognitive practice. Rather, they appear to redistribute the work students do: students engage less in open-ended solution generation and more in verification, diagnosis, and judgment. 
This redistribution is also visible in the homework data: students tended to receive higher scores on the GenAI-evaluation problems than on the corresponding direct-solution problems. However, in this implementation, this localized homework advantage did not translate into measurable differences in downstream exam performance.

Our findings have practical implications for the design of GenAI-based assignments. The absence of broad performance losses suggests that structured GenAI-evaluation activities can be incorporated into algorithms coursework without necessarily displacing all meaningful practice. However, the absence of measurable gains also suggests that simply asking students to grade an AI-generated solution may not be sufficient to improve transfer. This conclusion is supported by our survey results: the vast majority of students reported that the GenAI-based assignments did not change their study habits, 
but students who reported adapting their study behaviors rated the evaluation activity as significantly more helpful than those who did not. Future versions of such assignments may need to push students beyond simple diagnosis, for example by asking them to identify the first invalid step in an argument or algorithm, construct a counterexample where a proposed algorithm fails, propose a corrected algorithm or invariant, and explain why the corrected solution works. In this sense, the pedagogical value of GenAI evaluation may depend less on exposure to AI output itself and more on how carefully the evaluation task is scaffolded.

\subsection{Threats to Validity}\label{subsec:threats}

Several limitations should be considered when interpreting these results. First, although students were assigned to conditions using a randomized crossover design, randomization occurred at the homework-working-group level rather than at the individual-student level. This choice reflected the collaborative structure of the course and was intended to reduce contamination across conditions, but it also means that student outcomes within working groups may not be fully independent. Because clustering inflates Type I error rather than masking true effects, this concern bears asymmetrically on our findings: our null summative results would, if anything, be reinforced by accounting for within-group dependence, whereas our one positive finding (the localized homework difference) should be interpreted with corresponding caution as a result that may not survive a clustered analysis.

Second, the intervention compares two authentic instructional activities rather than isolating a single mechanism. Students in the GenAI-evaluation condition generated an LLM response, read and interpreted that response, evaluated its correctness, and explained any errors. Thus, the treatment should be understood as a structured GenAI-evaluation activity, not as a pure measure of the effect of exposure to AI-generated solutions or of grading alone.

Third, the GenAI-generated artifacts themselves may have varied across students. Although students were given standardized prompts and asked to use the same model, LLM outputs can differ across sessions and contexts. This variation is pedagogically realistic, but it may also introduce noise into the intervention. Instructors who use similar activities may need to decide whether to let students generate their own AI outputs or instead provide a common instructor-generated artifact for all students to evaluate.

Fourth, the crossover design took place within a live course in which topics changed over time. The first half of the course covered different material from the second half, and later topics were generally more difficult. As a result, differences between first-half and second-half outcomes may reflect topic difficulty, student maturation, accumulated course experience, or exam timing in addition to the instructional condition itself.

Fifth, our measures capture only some forms of learning. Exam scores and structurally aligned exam questions provide meaningful evidence about summative performance and transfer of algorithmic reasoning, but they may not capture all competencies targeted by GenAI-evaluation tasks, such as trust calibration, debugging AI-generated explanations, or recognizing subtle flaws in plausible-looking solutions. Conversely, self-reported helpfulness and study behavior are subjective and may be affected by recall, social desirability, or students' differing interpretations of the survey items.

Finally, the absence of statistically significant differences should not be interpreted as proof that the two activities are equivalent. Our results indicate that we did not observe evidence of meaningful performance differences in this course implementation. Establishing equivalence or non-inferiority would require an analysis designed specifically for that purpose.

\section{Conclusion and Future Work}

As GenAI tools become increasingly integrated into students' programming workflows, computing education faces a practical design question: how should coursework adapt when students can readily generate plausible solutions to traditional assignments? This study examined one response: replacing some direct solution construction with structured evaluation of GenAI-generated solutions in a junior-level algorithms course. Using a randomized crossover design, we found no evidence that this intervention improved summative performance or transfer to structurally aligned exam problems relative to traditional problem solving, but also no evidence of broad performance losses. Together, these findings suggest that structured GenAI-evaluation activities may be feasible components of algorithms instruction, but that their value depends on careful task design rather than on the mere presence of AI-generated content.

Future work should investigate which forms of scaffolding make GenAI-evaluation activities most effective. In particular, it would be useful to compare evaluation-only tasks with evaluation-and-repair tasks, to contrast student-generated AI outputs with standardized instructor-provided outputs, and to study whether these activities benefit some students more than others. More broadly, future research should examine how evaluation-centered GenAI activities affect skills that are increasingly important in AI-rich programming environments, including verification, critique, explanation, repair, and calibrated reliance on generated solutions, as well as the development of validated assessment instruments to measure these specific competencies.


\bibliographystyle{ACM-Reference-Format}
\bibliography{refs}

\end{document}